\documentclass[prd,twocolumn,showpacs,amsmath,amssymb,superscriptaddress,floatfix,nofootinbib]{revtex4}
\usepackage{mathrsfs,bm}
\usepackage{longtable,lscape}
\usepackage{txfonts}
\usepackage{amssymb}
\usepackage{indentfirst}
\usepackage{graphicx,,booktabs}
\usepackage{multirow}
\usepackage{color}
\usepackage{amssymb}

\begin{document}
\title{Pion-induced production of the $Z_c(3900)$ off a nuclear target}

\author{Yin Huang}
\email{huangy2014@lzu.cn}
\affiliation{Research Center for Hadron and CSR Physics, Lanzhou University and Institute
of Modern Physics of CAS, Lanzhou 730000,China}
\affiliation{School of Nuclear Science and Technology,
Lanzhou University, Lanzhou 730000, China}
\affiliation{Institute of modern physics,
Chinese Academy of Sciences, Lanzhou 730000, China}

\author{Jun He\footnote{Corresponding author}}
\email{junhe@impcas.ac.cn}
\affiliation{Research Center for Hadron and CSR Physics, Lanzhou University and Institute
of Modern Physics of CAS, Lanzhou 730000,China}
\affiliation{Institute of modern physics,
Chinese Academy of Sciences, Lanzhou 730000, China}
\affiliation{State Key Laboratory of Theoretical Physics,
Institute of Theoretical Physics, Chinese Academy of Sciences}

\author{Xiang Liu}
\affiliation{Research Center for Hadron and CSR Physics, Lanzhou University and Institute
of Modern Physics of CAS, Lanzhou 730000,China}
\affiliation{School of Physical Science and Technology, Lanzhou University, Lanzhou 730000, China}

\author{Hong Fei Zhang}
  \affiliation{School of Nuclear
Science and Technology, Lanzhou University, Lanzhou 730000, China}\affiliation{Institute of modern physics, Chinese Academy of
Sciences, Lanzhou 730000, China}

\author{Ju Jun Xie}
\affiliation{Research Center for Hadron and CSR Physics, Lanzhou University and Institute
of Modern Physics of CAS, Lanzhou 730000,China}
\affiliation{Institute of modern physics,
Chinese Academy of Sciences, Lanzhou 730000, China}
\affiliation{State Key Laboratory of Theoretical Physics,
Institute of Theoretical Physics, Chinese Academy of Sciences,Beijing 100190, China}

\author{Xu Rong Chen}
\affiliation{Research Center for Hadron and CSR Physics, Lanzhou University and Institute
of Modern Physics of CAS, Lanzhou 730000,China}
 \affiliation{Institute of modern physics, Chinese Academy of Sciences, Lanzhou 730000, China}

\date{\today}
\begin{abstract}

We investigate the possibility to study the charmoniumlike state
$Z_c(3900)$ through the pion-induced production off a nuclear target.  By
using a high-energy pion beam, the $Z_c(3900)$ can be produced off a proton or
nucleus though the Primakoff effect. The production amplitude is
calculated in an effective Lagrangian approach combined with the vector
dominance model.  The total cross sections of the $p(\pi^-, Z^-_c(3900))$ and $p(\pi^-, Z^-_c(3900)\to J/\psi\pi^-)$
reactions are calculated, and their order of magnitude is about 0.1 and 0.01nb, respectively,  with an assumption of branch ratio 10\% for the $Z_c(3900)$ decay in $J/\psi\pi$ channel. If the proton target is replaced by
a nuclear target, the production of the $Z_c(3900)$ enhances obviously.
The predicted total cross sections for the $A(\pi^-, Z^-_c(3900))$ and
$A(\pi^-, Z^-_c(3900)\to J/\psi\pi^-)$ reactions with $A=^{12}$C or $^{208}$Pb are on the order of magnitude of 100 and 10 nb, respectively, which is about one thousand times larger
than the cross sections off a proton target.  Based on these  the results, we
suggest the experimental study of the $Z_c(3900)$ by using high-energy
pion beams with a nuclear target at facilities such as COMPASS and
J-PARC.

\end{abstract}


\maketitle
\section{INTRODUCTION}

The resonant structure
$Z_c^\pm(3900)$ was reported recently by the BESIII and Belle Collaborations in the $J/\psi{}\pi^{\pm}$ invariant mass spectrum
through the electron-positron collision process $e^{+}e^{-}\to{}J/\psi{}\pi^{+}\pi^{-}$ at
$\sqrt{s}=4.26$ GeV~\cite{Ablikim:2013mio,Liu:2013dau} and later in an analysis of the CLEO-c
data~\cite{Xiao:2013iha}. A structure at $M=3883.9\pm1.5\pm4.2$
MeV was also observed in the $D\bar{D}^*$  invariant mass spectrum
in process $e^+e^-\to\pi^\pm
(D\bar{D}^*)^\mp$~\cite{Ablikim:2013emm}. Recently, a new neutral
state with mass $3894.8\pm2.3\pm3.2$ MeV was observed in
process $e^+e^-\to\pi^0\pi^0 J/\psi$ and interpreted as a neutral partner of
the $Z^\pm_c(3900)$~\cite{Ablikim:2015tbp}.

Although such a resonant structure has been  established in the experiment,
its origin is still in debate. Since the $Z_c^\pm(3900)$ carries a charge,
it cannot be put into a conventional quark model scheme as a
$c\bar{c}$ state.  Due to closeness to the $D\bar{D}^*$ threshold,
the molecular state interpretation of the $Z_c(3900)$ was proposed soon after
its observation~\cite{Wang:2013cya,Wilbring:2013cha,He:2014nya}. However, the
lattice QCD simulation suggested that a shallow bound state related to
the $Z_c(3900)$ cannot be formed in the $D\bar{D}^*$
interaction~\cite{Chen:2014afa,Lee:2014uta,Prelovsek:2013xba}.  In
Ref. \cite{He:2015mja}, the author suggested that the $Z_c(3900)$ should
be interpreted as a resonance instead of a bound state. The tetraquark
state interpretation was also proposed in
Refs.~\cite{Braaten:2013boa,Dias:2013xfa}. There exist many
nonresonant explanations of the $Z_c(3900)$ structure coming from some
special kinematics instead of from a genuine resonance. In
Ref.~\cite{Chen:2013coa}, the $J/\psi\pi$ invariant mass spectrum can
be reproduced through an initial-single-pion-emission mechanism.  The
cusp effect from triangle singularity was also adopted to explain the
$Z_c(3900)$ structure~\cite{Liu:2013vfa}.

More experimental data will provide more opportunity to understand
the structure of the so-called $XYZ$ particles. Most of our experimental knowledge
about $XYZ$ particles is from the electron-positron collision or the
$B$ decays. In the literature, other ways to observe the $XYZ$ particles
have been proposed. Experimental research for the charmoniumlike states $Z_c(4430)$
and $Y(3940)$ and the hidden-charm nucleon resonance in photoproduction have been proposed in Refs.~\cite{prd9,prd10,Huang:2013mua}.  In Ref.~\cite{prd11}, the cross section of
the $Z_c(3900)$ photoproduction off a proton was predicted to reach a
maximum value of 50 to 100 nb at $\sqrt{s_{\gamma N}}\sim10$ GeV. This
prediction inspired an experiment at  COMPASS, where photo-nucleon
energies cover the range $\sqrt{s_{\gamma{}N}}$ from 7 to 19 GeV.
Unexpectedly, no signal of the $Z_c(3900)$ was observed in the $J/\psi \pi^\pm$ mass spectrum~\cite{prd12}.  It seems that the decay channel
$Z_c(3900)^{\pm}\to J/\psi{}\pi^{\pm}$ cannot be the dominant one.
Another possible explanation about the COMPASS result is that the theoretically
predicted $Z_c(3900)$ photoproduction off a proton is
overestimated.  It is well known that the replacement of the  proton target by the nuclear target
will enhance the meson production~\cite{Badier:1983dg,Kaskulov:2011ab}. The nuclear target may be more appropriate to use to  produce the $XYZ$ particles.

The high-energy pion beams are available in facilities such as J-PARC and COMPASS. It is interesting to make a theoretical prediction about the pion-induced production to evaluate the feasibility to observe the $Z_c(3900)$ in experiment with a nuclear target. In this work, we focus on the charged pion ($\pi^-$)-induced production of $Z^-_c(3900)$. Empirically, the $s$-channel and $u$-channel contributions are much smaller than $t$-channel contribution especially at high energies ~\cite{prd9,prd10,Huang:2013mua}. The known decay channel of the $Z_c(3900)$ with pions is $Z_c(3900)\to J/\psi\pi$, so we take the  $J/\psi$ as the exchanged meson in the  $t$ channel. Since the $J/\psi NN$ vertex is OZI suppressed, the Primakoff effect will be introduced here.  As is well known in the photon-induced Primakoff effect, the high-energy pion beam will interact with the nucleon or nuclei through exchanging  a virtual photon.
Combined with the vector meson dominance (VMD) mechanism, the $Z_c(3900)$ can be produced.
In this work, we will study the $Z_c(3900)$ in the $A(\pi^-, Z_c^-(3900))$ and
$A(\pi^-, Z_c^-(3900)\to J/\psi\pi^-)$ reactions. Here $A=^{12}$C and $A=^{208}$Pb are taken as the examples for light and heavy nuclei, respectively.

This paper is organized as follows. In the next section, the $Z_c(3900)$ production off a proton is presented. Then we replace the proton by nuclei, and the cross sections for the $A(\pi^-, Z_c^-(3900))$ and
$A(\pi^-, Z_c^-(3900)\to J/\psi\pi^-)$ reactions  with $A=^{12}$C or $^{208}$Pb are calculated in Sec. III.   A summary is given in the last section.

\section{Pion-induced ${Z_c(3900)}$ production off a proton target}

\begin{figure}[h!]
\begin{center}
\includegraphics[height=0.46\textheight]{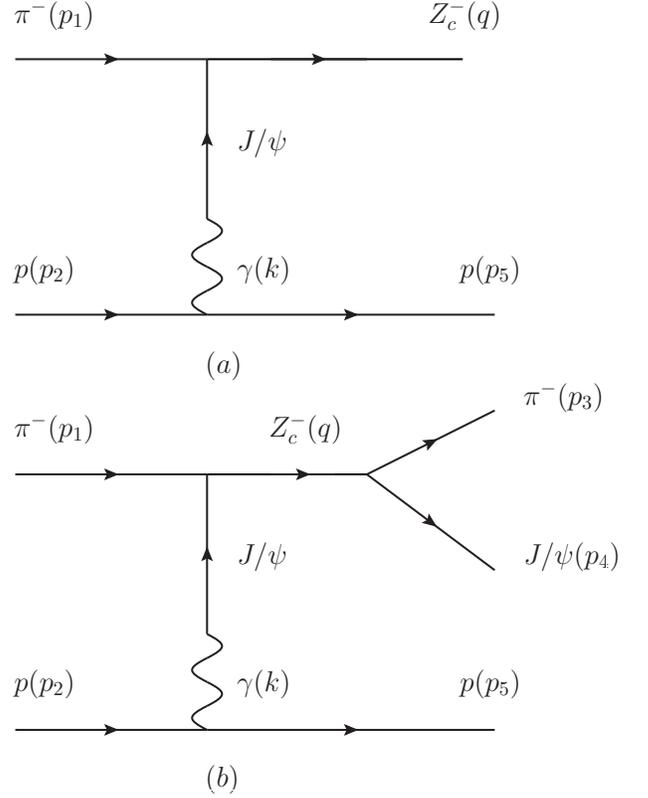}
\caption{Feynman diagrams for the $p({\pi^{-}}, {Z_c^{-}(3900)}) $ [upper figure (a)]and $p(\pi^-, Z^-_c(3900)\to J/\psi\pi^-)$ reactions [lower figure (b)].}\label{feydiagrams}
\end{center}
\end{figure}

The production mechanism through a pion-induced Primakoff effect is illustrated in Fig.~\ref{feydiagrams}. The pion interacts with the proton through a virtual photon.  Through the VMD mechanism, the virtual photon becomes a $J/\psi$ meson, which can interact with a pion to produce a $Z_c(3900)$ because the $Z_c(3900)$ was observed  first just in the $J/\psi\pi$ channel.  The cascade decay of the $Z_c(3900)$ to $J/\psi \pi$ will be also considered in this work [see Fig.~\ref{feydiagrams}(b)].

In the VMD mechanism, the
Lagrangian depicting the coupling of the intermediate state
$J/\psi$ with the photon is written as
\begin{equation}
{\cal{L}}_{J/\psi\gamma}=-\frac{eM_{J/\psi}^2}{f_{J/\psi}}V_{\mu}A^{\mu},
\end{equation}
where  $V$ and $A$ are the $J/\psi$ and photon fileds, $e$ is the unit charge, and $M_{J/\psi}$ and $f_{J/\psi}$ denote the mass and the decay constant of the vector meson $J/\psi$.
With the decay width $\Gamma_{J/\psi\to e^{+}e^{-}}=5.55\pm0.14\pm0.02$ KeV,
one obtains the parameter $e/f_{J/\psi}=0.027$.
A form factor is added in this vertex of a from
$F_{J/\psi\gamma}(k^2)={\Lambda^2}/(\Lambda^2-k^2)$
with cutoff $\Lambda$ choosing as the mass of the vector meson $J/\psi$£¬ as in Ref. \cite{Zhao:2007mfa} .

The coupling between photon and nucleon is depicted by an effective Lagrangian,
\begin{equation}
{\cal{L}}_{\gamma{}pp}=-e\bar{N}[\gamma^{\mu}A_{\mu}F_1(k^2)-\frac{\kappa_p}{2M_N}\sigma_{\mu\nu}\partial^{\nu}A^{\mu}F_2(k^2)]N,
\end{equation}
where $A$ and $N$ denote photon and proton field with mass $M_N$, respectively.
The anomalous magnetic momentum of the proton is $\kappa_p=1.97$, and the antisymmetric tensor is defined as $\sigma_{\mu\nu}=\frac{i}{2}(\gamma_{\mu\nu}-\gamma_{\nu\mu})$. The proton form factor $F_{1,2}(k^2)$ can be rewritten as electromagnetic form factors,
\begin{align}
&G_E(k^2)=F_1(k^2)+\frac{k^2}{4M_N^2}F_2(k^2),\nonumber\\
&G_M(k^2)=F_1(k^2)+F_2(k^2).
\end{align}
The experimental electromagnetic form factors  of the proton can be approximately described
by a dipole fit \cite{prd18},
\begin{align}
G_E^p(Q^2)=G_M^p(Q^2)/\mu_p=(1+{Q^2}/{m_D^2})^{-2},
\end{align}
where $m_D^2=0.71$ GeV$^2$, $Q^2=-k^2$, and $\mu_p=2.793$.

Here, we adopt an assignment of the spin-parity quantum number of the $Z_c(3900)$
as $J^{P}=1^{+}$ \cite{prd15}.   Under such assignment, the effective Lagrangian describing the $Z_c(3900)J/\psi\pi$ coupling is of the form ~\cite{prd9}
\begin{equation}
{\cal{L}}_{\pi\psi{}Z_c}=\frac{g_{\pi\psi{}Z_c}}{M_{Z_c}}(\partial_{\mu}\psi_{\nu}\partial^{\mu}\pi{}Z_{c}^{\nu}-\partial_{\mu}\psi_{\nu}\partial^{\nu}\pi{}Z_{c}^{\mu}),
\end{equation}
where $Z_c$ and $\psi$ denote the $Z(3900)$ field with mass $M_{Z_c}$ and the $J/\psi$ field, respectively. The coupling constant $g_{\pi\psi{}Z_c}$ can be determined by the decay
width of the $Z^{-}_c(3900)$ in ${}J/\psi\pi^{-}$ channel~\cite{prd15}. With an assumption that the branch ratio of the $Z_c(3900)$ in $J/\psi\pi$ is 100\%, we
have ${g_{\pi\psi{}Z_c}}/{M_Z}=1.50$ and $1.90$
for typical values $\Gamma_{Z^{-}_c(3900)\to{}J/\psi\pi^{-}}=29$~\cite{Ablikim:2013emm}
and $46$ MeV~\cite{Ablikim:2013mio}, respectively.

The form factor should be introduced to reflect the internal structure of the hadrons.
A form factor is introduced as
\begin{align}
F_{Z_c\psi\pi}(k^2)=\frac{\Lambda^2-M_{J/\psi}^2}{\Lambda^2-k^2}
\end{align}
for the exchanging $J/\psi$, and  a form factor is introduced as
\begin{align}
F_{Z_c\psi\pi}(q)=\frac{\Lambda^2-M_{Z_c}^2}{\Lambda^2-q^2}
\end{align}
for the vertex of the $Z_c^{-}(3900)$ decay to $J/\psi\pi^{-}$ [see Fig.~\ref{feydiagrams}(b)].
Empirically, the cutoff $\Lambda$ is close to 1 GeV. In this work we will discuss it with different values not far from 1 GeV.

\begin{figure}[h!]
\begin{center}
\includegraphics[height=0.31\textheight]{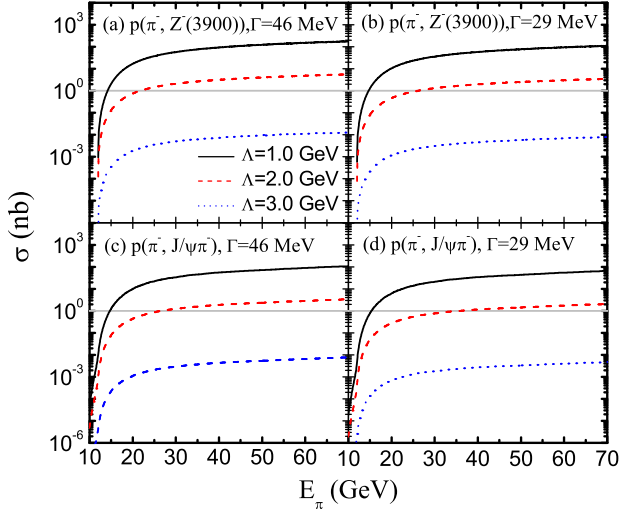}
\caption{(Color online) The total cross sections for the $p(\pi^-, Z_c^-(3900))$  reaction (panels a and b) and
$p(\pi^-, Z_c^-(3900)\to J/\psi\pi^-)$ reaction (panels c and d)
versus the $\pi^{-}$-beam energy $E_\pi$ with typical values of cutoff $\Lambda=$ 1 (full line), 2 (dashed line) and 3 (dotted line) GeV.
The two left panels a and c and two right panels b and d correspond to results taking $\Gamma_{Z_c^{-}\to{}J/\psi\pi^{-}}=46$ and $29$ MeV, respectively. The grey horizontal line is for 1 nb.
}\label{Crosssection2}
\end{center}
\end{figure}

With the above preparation, the cross sections for the $p(\pi^-, Z^-(3900))$ and
$p(\pi^-, Z^-(3900)\to J/\psi\pi^-)$ reactions  with variation of beam energies $E_{\pi}$ are presented
in  Fig.~\ref{Crosssection2}.
The total cross section of the $p(\pi^-, Z_c^-(3900))$ reaction increases rapidly near the threshold and becomes relatively stable at about $E_\pi=20$ GeV. The results with cutoff $\Lambda$=1, 2, and 3 GeV are presented in  Fig.~\ref{Crosssection2}, which suggests that the cross section decreases with the increase of the cutoff $\Lambda$ from 1 to 3 GeV. With pion beam energy larger than 20 GeV, the production cross sections with two choices of the decay widths, $\Gamma_{Z_c^{-}\to{}J/\psi\pi^{-}}=46$  and $29$ MeV, are on the order of magnitude of 1 nb  with cutoff $\Lambda$ in a range of $1\thicksim2$ GeV.

The cross section for the $p(\pi^-, Z^-_c(3900)\to J/\psi\pi^-)$ reaction is also presented in Fig.~\ref{Crosssection2} and a similar result is found as the $p(\pi^-, Z_c^-(3900))$ reaction because, that in the calculation, we assume that the branch ratio of the $Z_c(3900)\to J/\psi \pi$ channel is 100\%. The branch ratio has  not been determined in the experiment. If the physical branch ratio is not 100\%, the cross section for the $p(\pi^-, Z_c^-(3900))$  and
$p(\pi^-, Z^-_c(3900)\to J/\psi\pi^-)$ reactions should be multiplied by the branch ratio and the square of the branch ratio, respectively.

Hence, if we assume the branch ratio is  10\%, a reasonable estimation about the order of magnitude of the cross section of the pion-induced $Z_c(3900)$ production off a proton is 0.1 nb. If we observe it in the $J/\psi\pi$ channel, the cross section will be suppressed further by a order of magnitude, but it is difficult to detect  in the existing facilities.

\section{Pion-induced ${Z_c(3900)}$ production off a nuclear target}

For pion-induced ${Z_c(3900)}$ production off a nuclear target, the proton in Fig.~\ref{feydiagrams} should be replaced by a nucleus.
It follows that the interaction of the photon with a nucleus can be described by the Lagrangian~\cite{Kaskulov:2011ab},
\begin{align}
{\cal{L}}_{\gamma{}A}=-iG_{\gamma{}A}[C^{\dagger}(z)&\partial_{\mu}C(z)-(\partial_{\mu}C^{+}(z))C(z)]\nonumber\\
                     &\times\int{}d^{4}yF_{A}(y-z){\cal{V}}_{\gamma}^{\mu}(y)\label{eq9},
\end{align}
where the coupling constant $G_{\gamma{}A}=|e|Z$ with $Z$ being the number of protons in the nucleus,  $x,y,z$ denote
the three spatial components and $r$ is distant from the center of the nucleus. $C^{\dagger}$($C$) creates (annihilates) a nucleus with a definition as,
\begin{equation}
C^{\dagger}(z)=\int{}\frac{d^3\vec{k}^{'}}{\sqrt{(2\pi)^3}}\frac{\alpha^{\dagger}_{k^{'}}}{\sqrt{2k^{'0}}}e^{ik^{'}z},
\end{equation}
with  $k^{'}$, $\vec{k}^{'}$, and $k^{'0}$ being the four-momentum, three-momentum and energy, respectively.

The profile function $F_{A}$ of a nucleus makes the interaction of the photon field ${\cal{V}}_{\gamma}^{\mu}(y)$ nonlocal.
The profile function $F_{A}$ is normalized as
$\int{}d^4yF_{A}(y)=1$.
After the Fourier transformation, $F_A$ becomes a form factor of the nucleus
as a function of momentum-transfer
$q$~\cite{prd16},
\begin{align}
{\cal{F}}(k)=\frac{4\pi}{Z}\int{}\rho(r)_{ch}j_{0}(kr)r^2dr,\label{Eq: NFF}
\end{align}
where  $j_{0}(qr)=\sin(qr)/qr$ and $\rho(r)_{ch}$ is the ground-state charge density, which gives the
most direct physical insight into the distribution of protons inside the nucleus.
The charge-density normalization is given by
$4\pi{}\int{}\rho(r)_{ch}r^2dr=Z$,
 and the Woods-Saxon form
$\rho_{ch}(r)=\rho_0/(1+e^{(r-R)/d})$
is adopted in this work with charge radius $R =1.1 A^{1/3}$ fm and nuclear surface thickness $d = 0.53$ fm ~\cite{prd17}.
$\rho^{0}$ is a normalized coefficient, and $A$ is the mass number of nuclear.

The amplitude describing the reaction $A(\pi^{-}, Z_c^{-}(3900))$ is given by
\begin{align}
-iM^{\lambda}&=-i\int{}d^4x\int{}d^4y\int{}d^4z\int\frac{d^4k}{(2\pi)^4}F_{A}(y-z)\nonumber\\
             &\times{}(-\frac{e|e|Z M_{J/\psi}^2g_{Z\psi\pi}}{M_Zf_{J/\psi}})\frac{(p_2+p_5)^{\mu}}{\sqrt{2p_2^0}\sqrt{2p_{5}^0}}\frac{(g_{\mu\nu}-\frac{k_{\mu}k_{\nu}}{M_{J/\psi}^2})}{k^2(k^2-M_{J/\psi}^2)}\nonumber\\
             &\times[k\cdot{}p_1g^{\nu\alpha}-k^{\nu}p_1^\alpha]\epsilon^{\lambda}_{Z_c\alpha}(q)F_{Z\psi\pi}(k^2)F_{J/\psi\gamma}(k^2)\nonumber\\
             &\times{}e^{-i(p_2-p_5)z}e^{-ik(x-y)}e^{-i(p_1-q)x}\chi_P^{(-)*}(\vec{x}-\vec{z}),
\end{align}
where $\epsilon^{\lambda}_{Z_c}$ is the polarization vector of the $Z_c(3900)$, with $\lambda$ being its helicity.
If the $\pi^{-}J/\psi\to{}Z_c^{-}(3900)$ conversion occurs inside the nucleus, the outgoing meson wave acquires
an additional eikonal phase $\chi_P^{(-)*}$.  At high energies the distortion factor is well described by an eikonal form as
\begin{align}
\chi_p^{(-)*}(\vec{\zeta})=\exp[-\frac{1}{2}\sigma_{\pi}\int^{z}_{-\infty}dz^{'}\rho_{ch}(\vec{b},\hat{n}z^{'})],
\end{align}
where $\vec{\zeta}=\vec{x}-\vec{z}$, $\hat{n}$ is a unit three-vector in the direction of the outgoing meson, $b$ is the impact parameter and $\sigma_{\pi}=24.1$ mb stands for the $\pi{}p$ total cross section.

Analogously to the $Z_c(3900)$ production, the amplitude
for the $A(\pi^-, Z^-(3900)\to J/\psi\pi^-)$ reaction is

\begin{align}
-iM^{\lambda}&=-i\int{}d^4x\int{}d^4y\int{}d^4z\int\frac{d^4k}{(2\pi)^4}F_{A}(y-z)\nonumber\\
             &\times{}\left(-\frac{|e|e(M_{J/\psi}g_{Z\psi\pi})^2}{M_Z^2f_{J/\psi}}\right)\frac{(p_2+p_5)^{\mu}}{\sqrt{2p_2^0}\sqrt{2p_{5}^0}}\frac{(g_{\mu\nu}-\frac{k_{\mu}k_{\nu}}{M_{J/\psi}^2})}{k^2(k^2-M_{J/\psi}^2)}\nonumber\\
             &\times[k\cdot{}p_1g^{\nu\alpha}-k^{\nu}p_1^\alpha]\frac{-g_{\alpha\beta}+\frac{q_{\alpha}q_{\beta}}{M_{Z}^2}}{q^2-M_{Z}^2+iM_{Z}\Gamma_Z}F_{Z\psi\pi}(k^2)F_{J/\psi\gamma}(k^2)\nonumber\\
             &\times[p_3\cdot{}p_4g^{\eta\beta}-p_{3}^\eta p_4^\beta]\epsilon^{\lambda}_{J/\psi\eta}(p_4)F_{Z\psi\pi}(q^2)\nonumber\\
             &\times{}e^{-i(p_2-p_5)z}e^{-ik(x-y)}e^{-i(p_1-q)x}\chi_P^{(-)*}(\vec{x}-\vec{z}),
\end{align}
where $\epsilon^{\lambda}_{J/\psi}$ is the polarization vector of the $J/\psi$, and  $\Gamma_Z$ is total width of the $Z_c(3900)$.

\begin{figure}[h!]
\includegraphics[height=0.27\textheight]{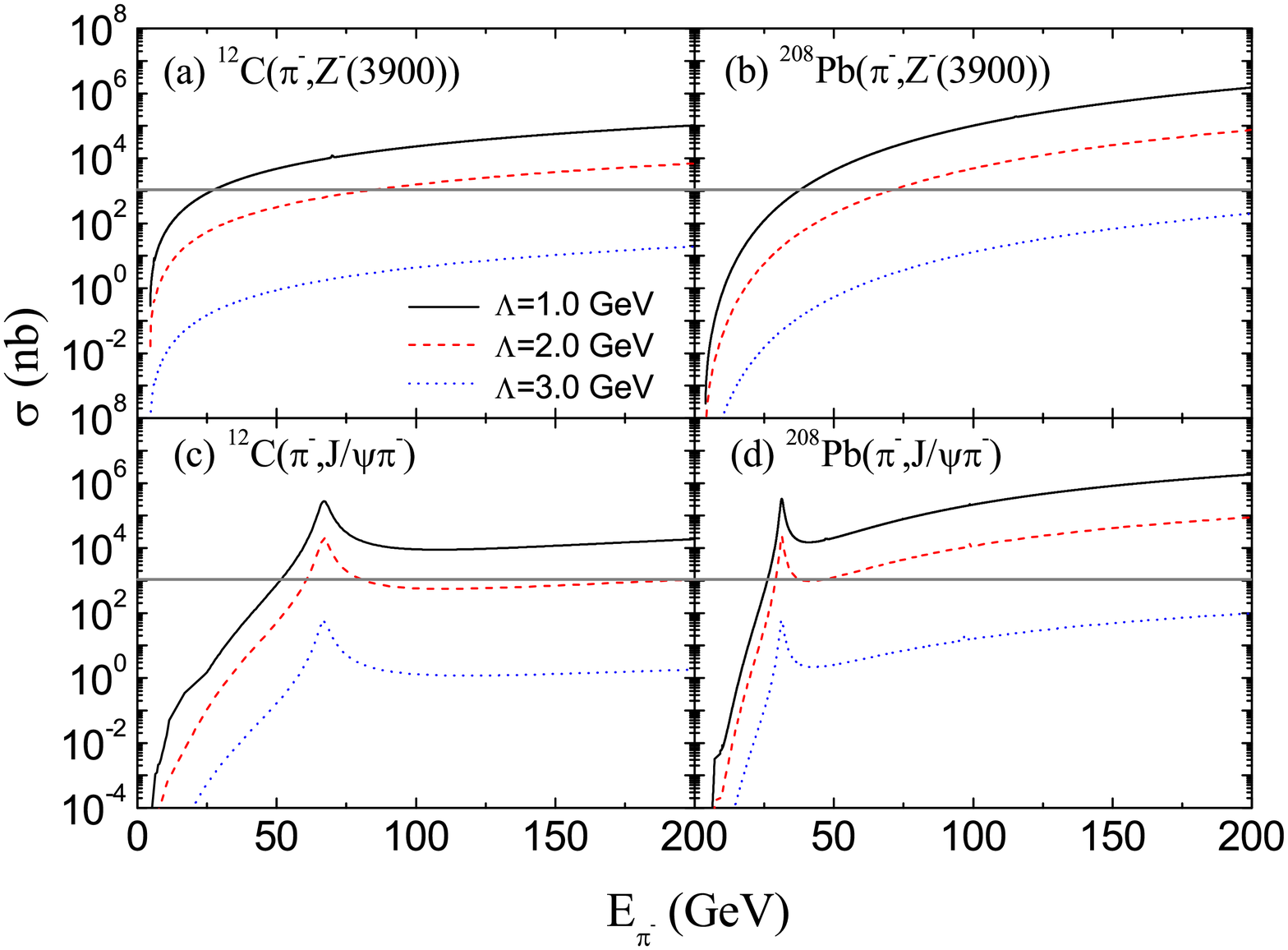}
\caption{(Color online) The total cross sections for the $A(\pi^-, Z_c^-(3900))$  reaction (panels a and b) and
$A(\pi^-, Z^-(3900)\to J/\psi\pi^-)$ reaction (panels c and d)
versus the $\pi^{-}$-beam energy $E_\pi$ with  $\Gamma=29$ MeV and $A=^{12}$C (panels a and c) and $^{208}$Pb (panels b and d).
The solid (black), dashed (red) and dotted (blue) lines correspond to results with $\Lambda=1$, 2 and 3 GeV, respectively. The grey horizontal line is for 1000 nb.
}\label{Crosssection4}
\end{figure}

With those amplitudes, the total cross sections for  the $A(\pi^-, Z^-(3900))$ and
$A(\pi^-, Z_c^-(3900)\to J/\psi\pi^-)$ reactions with  variation of the beam energies $E_{\pi}$ are calculated and presented
in  Fig.~\ref{Crosssection4}. Here, $A=^{12}$C and $^{208}$Pb are taken as the examples for light and heavy nuclei, respectively. Here we only consider $\Gamma=29$ MeV, because the cross sections with two choices of the decay width  are on the same order of magnitude  as the proton target.

The shapes of the cross sections of the $Z_c(3900)$ productions for both $A=^{12}$C and $^{208}$Pb are analogous to the proton case. Obvious enhancement of  the $Z_c(3900)$ production can be observed after replacing the proton target with the nuclear target [see Fig. 3(a) and Fig. 3(b)].
At energies higher than about 30 GeV, the cross sections with two nuclei are on the order of magnitude of 1000 nb or more with cutoff $\Lambda$ in a range of $1\thicksim2$ GeV, with which the cross section with the proton target is on the order of magnitude of $1$ nb (a line for 1000 nb is marked at Fig~\ref{Crosssection4}). A conservative estimation about order of magnitude of  the total cross section of the $A(\pi^-, Z_c^-(3900))$  reaction is 1000 nb if the beam energy is large enough. It is also found that the cross section for the heavier nuclei $^{208}$Pb is larger than that of the light nuclei $^{12}$C.

The cross section of the $p(\pi^-, Z_c^-(3900)\to J/\psi\pi^-)$ reaction  is almost the same as that of the  $p(\pi^-, Z_c^-(3900))$ reaction with 100\% branch ratio of the $Z_c(3900)$ decay in $J/\psi$ channel [see Fig.~3 (c) and Fig.~3 (d)]. However, different from the $A(\pi^-, Z_c^-(3900))$ reaction, the cross section of the $A(\pi^-, Z_c^-(3900)\to J/\psi\pi^-)$ interaction has a peak at energy about 70 or 30 GeV for $A=^{12}$C or $^{208}$Pb, respectively.
As  the $A(\pi^-, Z_c^-(3900))$ interaction, the cross section for the heavier nuclei $^{208}$Pb is larger than that of the light nuclei $^{12}$C. At higher energies, the cross section in $J/\psi\pi$ channel for $^{12}$C is smaller than that of the direct $Z_c(3900)$  production while the cross section in the $J/\psi\pi$ channel for $^{208}$Pb is close to that of the direct $Z_c(3900)$  production.

It is reasonable to estimate that the $Z_c(3900)$ production off a nuclear target is on the order of magnitude of 100 nb, based on the calculation in this work with an assumption of branch ratio of the $Z_c(3900)$ decay in the $J/\psi\pi$  channel as 10\%. If observed in the $J/\psi\pi$ channel, the cross section is on the order of magnitude of 10 nb.

\section{SUMMARY}

In this work, the $Z_c(3900)$ production through a pion-induced Primakoff effect combined with the VMD mechanism is studied with the effective Lagrangian method. The cross sections for the $A(\pi^-, Z_c^-(3900))$ and
$A(\pi^-, Z^-(3900)\to J/\psi\pi^-)$  reactions are calculated with $A$ being a proton, a $^{12}$C or a $^{208}$Pb.

The cross section of pion-induced $Z_c(3900)$ production off a proton is on the order of magnitude of 0.1 nb, which is difficult to  detect in the existing facilities. After replacing the proton with the nuclei, the total cross section of the $Z_c(3900)$ production enhances obviously to an order of magnitude of 100 nb  with an assumption of the branch ratio as 10\% for the $Z_c(3900)$ in the $J/\psi$ channel. Under the same branch ratio assumption , the cross section of the $A(\pi^-, Z^-(3900)\to J/\psi\pi^-)$  reaction is suppressed by 1 order of magnitude  to 10 nb due to an additional $Z_c(3900)\to J/\psi\pi$ vertex involved.  Based on the results, we
suggest the experimental study of the $Z(3900)$ by using high-energy
pion beams with nuclear targets at facilities such as COMPASS and
J-PARC.

The enhancement of the meson production with a nuclear target compared with a proton target is not limited to the $Z_c(3900)$ considered in the current work, which should exist in the production of other XYZ particles. In the literature, predictions of photon- or pion-induced productions of XYZ particles with a  proton target have been suggested by many authors ~\cite{prd9,prd10,prd11}. Based on the predictions in this work, we suggest studying  the XYZ particle production with a nuclear target rather than a proton target.

\section*{Acknowledgments}
We would like to thank Alexey Guskov
for suggesting that we study the $Z_c(3900)$ through the pion-induced Primakoff effect and for comments on the manuscript.  This project was partially supported by the Major
State Basic Research Development Program in China (No.
2014CB845400) and the National Natural Science Foundation of
China (Grants No. 11275235, No. 11035006, No.11175220,No. 11175073, and No. 11222547).

\end{document}